\documentclass[lettersize,journal]{IEEEtran}
\usepackage{amsmath,amsfonts}
\usepackage{algorithmic}
\usepackage{algorithm}
\usepackage{array}
\usepackage[caption=false,font=normalsize,labelfont=sf,textfont=sf]{subfig}
\usepackage{textcomp}
\usepackage{stfloats}
\usepackage{url}
\usepackage{verbatim}
\usepackage{graphicx}
\usepackage{soul}
\usepackage{textcomp}
\usepackage{stfloats}
\usepackage{float}
\usepackage{url}
\usepackage{pgf}
\usepackage{verbatim}
\usepackage{graphicx}
\usepackage{cite}
\usepackage{amsmath}
\usepackage{tikz}
\usepackage{xcolor}
\usetikzlibrary{arrows.meta}
\usepackage{pgfplots}
\usepackage{standalone}
\usepackage{xcolor}
\usepackage{cite}
\usepackage{soul}
\usetikzlibrary{external}
\usepackage{xcolor}
\sethlcolor{yellow}  
\usepackage[font=footnotesize]{caption}
\usepackage{graphicx}
\usepackage[nolist]{acronym}
\usepackage{siunitx}
\DeclareMathOperator{\dB}{dB}

\DeclareMathOperator{\MHz}{MHz}
\DeclareMathOperator{\GHz}{GHz}

\DeclareMathOperator{\meters}{m}

\usepackage{titlesec}
\usepackage{multirow}
    
 \usepackage{xcolor,cite,etoolbox}
\makeatletter 
\pretocmd\@bibitem{\color{black}\csname keycolor#1\endcsname}{}{\fail}
\newcommand\citecolor[1]{\@namedef{keycolor#1}{\color{blue}}}
\makeatother

\usepackage{titlesec}

\titlespacing{\subsection}{3pt}{*0.05}{*0.1}
\titlespacing{\subsubsection}{3pt}{*0.05}{*0.1}
\begin{acronym}
	\acro{LoS}{line-of-sight}
    \acro{CSI}{channel state information}
	\acro{NLoS}{non-line-of-sight}
	\acro{InH}{indoor hotspot}
	\acro{InF}{indoor factory}
	\acro{UMi}{urban microcell}
	\acro{SHF}{super high frequency}
	\acro{DS}{delay spread}
	\acro{RMS}{root mean square}
	\acro{AS}{angular spread}
        \acro{DSS}{dynamic spectrum sharing}
	\acro{3GPP}{$3^{\text{rd}}$ Generation Partnership Program}
    \acro{NGJ-MB}{next generation jammer mid-band}
	\acro{RF}{radio frequency}
	\acro{PHY}{physical layer}
	\acro{ACLR}{adjacent channel leakage power ratio}
        \acro{FCC}{Federal Communications Commission}
        \acro{NTIA}{National Telecommunications and Information Administration}
         \acro{CFO}{carrier frequency offset}
         \acro{CRB}{Cram{\'{e}}r-Rao bound}
         \acro{AWGN}{additive white Gaussian noise}
         \acro{FFT}{fast Fourier transform}
         \acro{CP}{cyclic prefix}
         \acro{MSE}{mean squared error}
         \acro{MRC}{maximum ratio combining}
         \acro{CPE}{customer-premises equipment}
         \acro{TRP}{transmission and reception point}
         \acro{PSD}{power spectral density}
         \acro{SE}{spectral efficiency}
         \acro{AR}{augmented reality}
         \acro{XR}{extended reality}
         \acro{ISAC}{integrated sensing and communications}
         \acro{BS}{base station}
         \acro{FR1}{frequency range~1}
         \acro{FR2}{frequency range~2}
         \acro{FR3}{frequency range~3}
         \acro{PLE}{pathloss exponent}
         \acro{mmWave}{milli-meter wave}
         \acro{gNB}{gNodeB}
         \acro{NR}{new radio}
         \acro{UE}{user equipment}
         \acro{MC}{mutual coupling}
         \acro{SINR}{signal-to-interference-plus-noise ratio}
         \acro{ITU}{International Telecommunication Union}
         \acro{MIMO}{multiple-input and multiple-output}
         \acro{ADC}{analog-to-digital converter}
         \acro{DAC}{digital-to-analog converter}
         \acro{PA}{power amplifier}
         \acro{O-RU}{open radio unit}
         \acro{DoF}{degrees-of-freedom}
         \acro{UL}{uplink}
         \acro{DL}{downlink}
         \acro{SHO}{soft handover}
         \acro{CDMA}{code-division multiple access}
         \acro{VR}{virtual reality}
         \acro{QAM}{quadrature amplitude modulation}
         \acro{LTE}{long-Term Evolution}
         \acro{SNR}{signal-to-noise ratio}
         \acro{RMT}{random matrix theory}
         \acro{ELAA}{extremely large aperture array}
         \acro{NF}{near-field}
         \acro{DT}{digital twin}
         \acro{FF}{far-field}
         \acro{RT}{ray tracing}
         \acro{AI}{artificial intelligence}
         \acro{RAN}{radio access network}
         \acro{O-RAN}{open radio access network}
         \acro{RIC}{RAN intelligent controller}
         \acro{EC}{ergodic capacity}
         \acro{ULA}{uniform linear array}
         \acro{AoA}{angle-of-arrival}
         \acro{SAR}{synthetic aperture radar}
         \acro{RIS}{reconfigurable intelligent surface} 
         \acro{ILFD}{injection-locked frequency divider}
         \acro{TSPC}{true single-phase clock}
         \acro{ET}{extended target}
         \acro{UWB}{ultra-wide band}
         \acro{SnC}[S\&C]{sensing and communication}
         \acro{RnC}[R\&C]{radar and communication}
         \acro{PAS}{payload-aided sensing}
         \acro{SB}{spectrum block}
         \acro{RaaS}{Radar‑as‑a‑Service}
         \acro{MAC}{medium access control}
         \acro{mMIMO}{massive MIMO}
         \acro{NB}{narrowband}
         \acro{EIRP}{equivalent isotropic radiated power}
         \acro{TDoc}{technical document}
         \acro{RAN}{radio access network}
         \acro{WG1}{working group}
         \acro{CIF}{close-in free space model with a frequency-weighted path loss exponent}
         \acro{TSG}{technical specification group}
         \acro{ASA}{angular spread of arrival}
         \acro{DRT}{deterministic-random tradeoff}
         \acro{OTFS}{orthogonal time frequency space}
         \acro{AFDM}{affine frequency division multiplexing}
         \acro{REM}{radio environment map}
         \acro{MNO}{mobile network operator}
         \acro{AF}{ambiguity function}
         \acro{OFDM}{orthogonal frequency-division multiplexing}
         \acro{RMSE}{root mean square error}
         \acro{QoS}{quality-of-service}
         \acro{ACF}{auto-correlation function}
         \acro{TDD}{time division duplexing}
         \acro{GNSS}{global navigation satellite system}
         \acro{AF}{ambiguity function}
         \acro{PSK}{phase-shift keying}
\end{acronym}

\begin{document}

\title{From Coverage to Sensing: ISAC meets FR3}
\author{Ahmad Bazzi,
Florian Gast,
Fan Liu,
Shi Jin,
Gerhard Fettweis,
Marwa Chafii
\thanks{Ahmad Bazzi and Marwa Chafii are with the Engineering Division, New York University (NYU) Abu Dhabi, 129188, UAE and NYU WIRELESS,
NYU Tandon School of Engineering, Brooklyn, 11201, NY, USA
(email: ahmad.bazzi@nyu.edu, marwa.chafii@nyu.edu).}
\thanks{Florian Gast is with the Barkhausen Institut, Germany (email: florian.gast@tu-dresden.de).}
\thanks{Fan Liu and Shi Jin are with the School of Information Science and Engineering, Southeast University, Nanjing, China. (email: fan.liu@seu.edu.cn, jinshi@seu.edu.cn).}
\thanks{Gerhard Fettweis is with the Barkhausen Institut, Germany and Vodafone Chair TU Dresden, Centres
6G-life \& CeTI\& 5G++ Lab Germany \& SEMECO, Germany. (email: gerhard.fettweis@barkhauseninstitut.org).}
}

\markboth{Accepted by IEEE Communications Magazine, 2026}%
{Shell \MakeLowercase{\textit{et al.}}: A Sample Article Using IEEEtran.cls for IEEE Journals}

\maketitle

\begin{abstract}
Future 6G systems are expected to exploit upper midband spectrum in frequency range 3 (FR3) not only for high throughput communications, but also for sensing services such as localization, detection, and situational awareness. The following paper develops a concrete path from today's coverage-oriented deployments to FR3 networks that treat sensing as a native function. We first show how existing FR2 radars can be time-multiplexed and coordinated under a $6$G medium access control as radar-as-a-service, forming a bridge between legacy sensing and network-managed integrated sensing and communications (ISAC). We then propose a hierarchical FR3 beam-alignment strategy in which coarse access occurs at lower frequencies and refinement occurs at upper FR3, and quantify the resulting sensing and communication capabilities via range-angle Cram{\'{e}}r-Rao bounds in the near field. We identify intra- and inter-beam squint phenomena specific to wideband FR3 arrays, and discuss design approaches to mitigate them. On the signal-processing side, we argue that FR3 sensing cannot rely solely on pilot resources and discuss how much sensing information can be extracted from payload resource elements. We further highlight the role of calibrated FR3 channel simulators and real-time models as the core of wireless digital twins for training and evaluating ISAC algorithms, and discuss how massive MIMO and dense or distributed deployments at FR3 naturally act as large reconfigurable sensor arrays.
\end{abstract}

\section{Introduction}
\label{sec:intro}

\IEEEPARstart{A}{\lowercase{s}} \textcolor{black}{$6$G accelerates, the \ac{FR3} upper mid-band ($7$-$24 \GHz$) has emerged as a “Goldilocks” spectrum. It uniquely balances wide bandwidth and macro-cell coverage, avoiding the severe path loss of mmWave and the capacity limits of FR1. Regulators (e.g., 3GPP, FCC) increasingly view \ac{FR3} as essential for expanding \ac{AI}-driven services while coexisting with incumbent satellite and defense applications \cite{bazzi2025upper}.}

\textcolor{black}{Recent works establish the foundational elements of next-generation \ac{ISAC}, including \ac{AI}-enhanced processing \cite{10663823}, \textcolor{black}{\ac{NF}} wideband operations \cite{10663786}, and unified frameworks integrating sensing, communication, computation \cite{10812728}, and zero-energy \ac{RIS} powering \cite{10720877}. Concurrently, spectrum strategies position \ac{FR3} as a prime candidate for 6G, which balances favorable coverage with massive bandwidths \cite{bazzi2025upper}. As 3GPP Release 19 lays the standardization groundwork \cite{10918335}, system-level analyses demonstrate that adaptive, Open RAN-based networks can dynamically share FR3 spectrum and safely coexist with incumbents \cite{10459211, 10634041}. Furthermore, these deployment strategies are increasingly supported by robust propagation models, including \ac{RT} and frequency-dependent foliage-attenuation studies \cite{10993391}.}

\textcolor{black}{This article charts a path toward native \ac{FR3} sensing. We detail \ac{MAC}-layer \ac{RaaS} in \ac{FR2} and \ac{PHY}-layer innovations including hierarchical alignment, \ac{NF} gains, and allude to new intra- and inter-beam squints phenomena on \ac{FR3}. Finally, we evaluate \ac{SnC} coverage, \ac{PAS} design under \ac{DRT}, and the use of \acp{DT} and \ac{mMIMO} for ``network-as-a-sensor'' operations.}
\textcolor{black}{The primary audience includes wireless communication researchers\textcolor{black}{,} engineers, but also network operators, and $6$G system architects exploring \ac{ISAC}. Moreover, hardware designers can learn about the physical layer complexities of the FR3 spectrum, such as optimizing near field estimations and mitigating wideband beam-squint effects. Meanwhile, telecom operators and regulators will benefit from the pragmatic blueprint for coordinating legacy radars under a $6$G MAC to monetize sensing as a native network service.}

\section{\textcolor{black}{Background: ISAC \& FR3 Goldilocks Spectrum}}
\label{sec:background}
\subsection{The paradigm shift towards \ac{ISAC}}
Traditionally, wireless \ac{SnC} have operated in strict isolation, utilizing separate hardware, waveforms, and frequency bands. \ac{ISAC} intends to use a single hardware platform and a unified waveform to simultaneously transmit data and sense the physical environment. In a $6$G context, this means a cellular \ac{BS} does not merely beam data to a user's device but actively repurposes the scattered radio echoes to localize targets, track mobility, and map its surroundings.

\subsection{Physical Synergies of \ac{ISAC} and \ac{FR3}}
To make high-resolution \ac{ISAC} a reality without requiring impractically dense network deployments, the industry is turning to \ac{FR3}, which spans from $7.125$ GHz to $24.25$ GHz, that is uniquely positioned between the highly congested sub-6 GHz bands \ac{FR1} and the millimeter-wave bands \ac{FR2}.  Because \ac{FR3} \emph{combines the best of both \ac{FR1} and \ac{FR2} worlds}, it can balance the capabilities of good coverage while securing large bandwidths for sensing.

The strategic relevance of \ac{FR3} lies in its physical propagation characteristics. Radar resolution is fundamentally dictated by available bandwidth. Unlike the narrow channels available in \ac{FR1}, \ac{FR3} can provide wide, contiguous spectrum blocks (e.g., $100$ to $400$ MHz), which is the exact physical requirement necessary to achieve centimeter-level range accuracy for localization.
Simultaneously, FR3 wavelengths are physically large enough to experience less severe path loss, atmospheric absorption, and blockage than \ac{FR2} mmWave signals \cite{bazzi2025upper}, which allows \ac{FR3} signals to penetrate foliage and better diffract around urban obstacles. Also, the centimeter-scale wavelengths of FR3 are small enough to allow operators to pack \ac{mMIMO} arrays into compact form factors.

\section{From 77 GHz to 28 GHz: FR2 Radar for FR3 ISAC}
\label{sec:pragmatic-ISAC}


\begin{figure*}[t]
    \centering
    \includegraphics[width=0.95\textwidth]{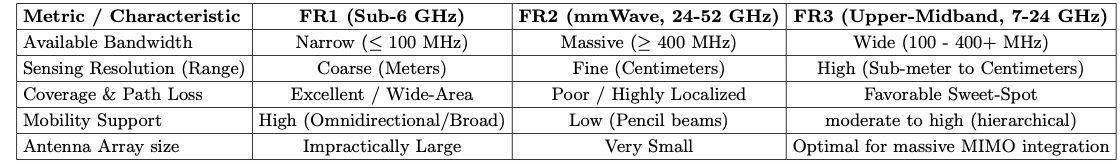} 
    
    \captionof{table}{\textcolor{black}{Comparative Analysis of Frequency Bands for ISAC}}
    \label{table:compare}
\end{figure*}

\subsection{MAC-Managed Radar \& RaaS on FR2}
\label{sec:mac-managed-radar}
\textcolor{black}{Coordinated scheduling of existing \ac{RnC} technologies offers a cost-efficient alternative to in-band \ac{ISAC} \cite{fettweis2025pragmatic}. The architecture avoids monostatic self-interference while preserving optimized, application-specific protocol stacks through managing \ac{FR2} radar sensors via \ac{FR1} or lower \ac{FR3} \ac{MAC} mechanisms.\cite{fettweis2025pragmatic}}

\begin{figure}
    \centering
    \includegraphics[width=1\linewidth]{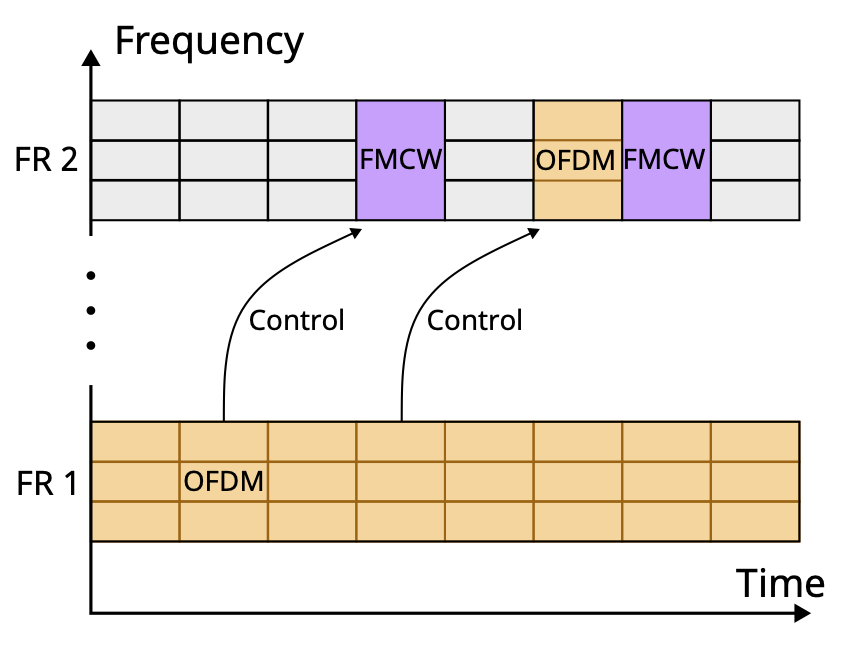}
    \caption{\textcolor{black}{Proposed pragmatic ISAC, where the currently employed OFDM in
FR1 controls sensing windows in FR2. If necessary, communications operation
can still be scheduled in FR2 using the MAC in FR1.}}
    \label{fig:PragmaticISAC}
\end{figure}

\textcolor{black}{\ac{FR2} is attractive for this purpose: it is underutilized, hotspot-centric, and provides bandwidth for high-fidelity sensing. To prevent mutual interference seen in the $77\,\text{GHz}$ band, precise scheduling of SnC windows is required. The Gearbox-PHY grid resource plane realizes this, coordinating heterogeneous SnC modes on shared resources across multiple frequency bands.}
A visualization of this idea is presented in Fig.~\ref{fig:PragmaticISAC}, where \ac{FR1} band \ac{MAC} is used to control \ac{SnC} windows in \ac{FR2}. 

\textcolor{black}{This transforms unmanaged $77 \GHz$ radars into schedulable and interference-controlled network functions with guaranteed \ac{SINR} and licensing compliance. Time-sliced sensing supports various mission profiles, while packaging scheduled windows and radar outputs as radio-environment-map data enables \ac{RaaS} operators to monetize sensing slices and datasets.}
\textcolor{black}{FR2-under-FR1/FR3 coordination supports mission profiles such as low-airspace safety, drone detection, and infrastructure monitoring. This synergy leverages FR3 for broad connectivity, while using FR2 for localized high-accuracy sensing for specialized verticals.}

\textcolor{black}{A practical implementation of the RaaS architecture, detailed in \cite{fettweis2025radiosensing}, requires a shared time reference across distributed FR2 radar nodes. Since the distributed radar nodes require connectivity for control and data aggregation, this can be provided either by the cellular network, which already maintains tight synchronization for \ac{TDD} operation, or by external references such as \ac{GNSS}. Using this common time base, radar transmissions can be organized into predefined sensing slots aligned with the cellular frame structure. These sensing windows could for example be scheduled via the control plane or preconfigured over the user plane for periodic sensing missions, enabling minimal radar-to-communication interference.}
\subsection{Downbanding 77\,GHz Radars to 26-28\,GHz}

The automotive radar stack is already mature and low‑cost at $24$ GHz and $77$ GHz and down‑banding those $77$ GHz MIMO radar chipsets to FR2 or upper FR3 is technically straightforward. The downbanding procedure can bring an existing ecosystem, which is already established for angular-chirp signaling with range-velocity processing, into a licensed band that mobile operators control. The motivation is that 77 GHz is congested, and many radars require more than $40-60$ dB of \ac{SINR} to function well. Therefore, relocating part of the workload to FR2 under a $6$G \ac{MAC} at the upper edge of FR3 lets the network allocate guarded $250-500$ MHz bandwidth of sensing slots with identification signaling and cross‑carrier triggers from FR1/FR3 control, turning self‑jamming into managed coexistence with predictable SINR, which can also leverage FR2 sensing that feeds \ac{RaaS} and \textcolor{black}{\ac{REM}} products.
State-of-the-art automotive radars operating at \SIrange{77}{81}{GHz} can utilize bandwidths of up to \(\SI{4}{GHz}\), achieving a resolution of roughly \(\SI{3.75}{cm}\).
By comparison, the \ac{FR2} bands $n257$ (\SIrange{26.5}{29.5}{GHz}, used, e.g., in the USA and Japan \cite{ericsson2021mmwave}) and $n258$ (\SIrange{24.25}{27.5}{GHz}, used, e.g., in the USA, Europe, and China \cite{ericsson2021mmwave}) can provide \(\SI{3}{GHz}\) and \(\SI{3.25}{GHz}\) of total bandwidth, respectively \cite{ETSI_138101_2}.
If a substantial portion of one of these spectra were aggregated, the theoretical resolution would reach about \( \SI{5}{cm}\), only slightly inferior to that of the \SI{77}{GHz} band. In practice, most \textcolor{black}{\acp{MNO}} operate with approximately \SI{800}{MHz} in the \ac{FR2} range \cite{ericsson2021mmwave}, which remains sufficient for long-range sensing. The effective radar bandwidth is often chosen smaller than the maximum available, as longer-range applications do not necessarily benefit from exploiting the full bandwidth allocation.
Importantly, leveraging such managed spectrum would allow operators to deploy radar in a controlled interference environment, thereby unlocking a wide range of new applications based on reliable and predictable sensing performance.
\textcolor{black}{We also provide a comparative analysis with FR1/FR2 \ac{ISAC} to position \ac{FR3} merits for \ac{ISAC} in Table \ref{table:compare}.}
\textcolor{black}{
FR3-assisted ISAC enables a pragmatic architecture in which cellular control schedules sensing windows for FR2 radar nodes. 
A practical path is to reuse mature 77 GHz FMCW radar concepts in the 26–28 GHz FR2 bands and coordinate distributed sensing via slot-based MAC scheduling.
Compared to fully integrated in-band ISAC, this approach sacrifices some communication capacity due to reserved sensing slots, but significantly lowers complexity and cost by keeping \ac{RnC} stacks separate.}

\section{Hierarchical Beam Alignment at FR3} 
\label{sec:beamalignment}
\begin{figure}[t!]
    \centering
    \includegraphics[width=0.5\textwidth]{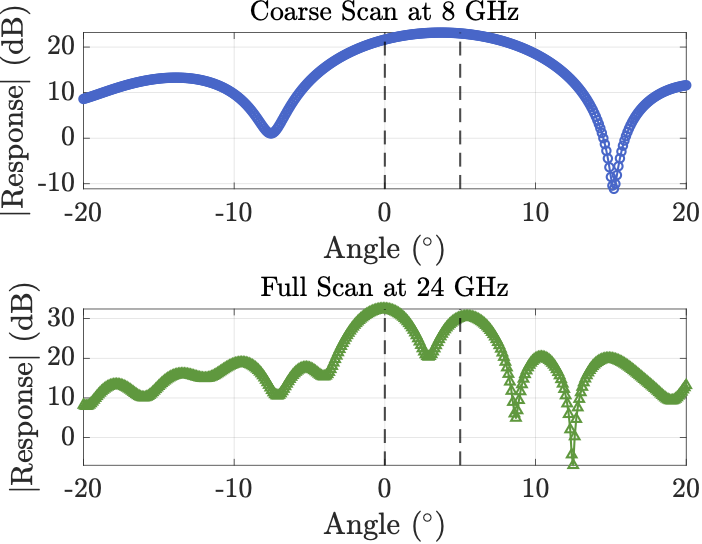}
    \caption{\textcolor{black}{Comparison of the beamforming response at two tiers. The top plot illustrates the coarse scan at $8\GHz$ with a $10$-element array, while the bottom panel shows the full scan at $24\GHz$ with a $30$-element array. Vertical dashed lines mark the two angles in the simulation ($0^\circ$ and $5^\circ$), highlighting how an initial low-frequency scan narrows the search range before a fine, high-frequency refinement.}}
    \label{fig:beamalignment}
\end{figure}
\textcolor{black}{At higher frequencies, \ac{FR3} hierarchical beam management optimizes user experience. Coarse, low-frequency wide beams establish initial directions robust to mobility and blockage. Subsequently, targeted upper-\ac{FR3} narrow-beam refinement confirms \ac{LoS} while minimizing overhead and interference. This two-stage design, integrated with Gearbox-PHY control and data planes, accelerates locking and provides the spatial precision required for sensing without exhaustive scanning.}
The outcome is a beam-management loop with coarse information preventing unnecessary scanning, while fine measurements deliver the spatial precision needed for sensing. 
\textcolor{black}{As shown in Fig. \ref{fig:beamalignment}, an $8 \GHz$ coarse scan narrows the angular search before a refined $24 \GHz$ scan within a $\pm 5^\circ$ window. The hierarchy balances $8 \GHz$ penetration tolerance with $24 \GHz$ alignment accuracy to optimize spectral efficiency and sensing.}
\textcolor{black}{Hierarchical design accommodates power-constrained $6$G devices like \ac{XR} glasses and sensor nodes, which cannot sustain exhaustive \ac{FR3} beam sweeps. Offloading search tasks to a lower-frequency stage minimizes sounding transmissions and computation, while delivering the narrow beams required for sensing and modulation, which then contributes to extended battery life and stabilizes initial access, especially in uplink-heavy scenarios.}

\textcolor{black}{It is important to note the deployment prerequisites for the frequency-spanning hierarchical beam alignment. Naturally, executing a coarse scan at 8 GHz and refining at 24 GHz requires the mobile network operator to possess non-contiguous frequency allocations that span the extremes of the upper-midband. In essence, the coarse scan at lower \ac{FR3} is suited in cases where high penetration losses are a concern, which can be combatted by the lower frequencies of \ac{FR3}, then a refinement of beam alignments can be attained at the higher \ac{FR3} frequencies, especially in line-of-sight cases.}
\textcolor{black}{The \ac{FR3} enables  hierarchical beam alignment that uses low-frequency wide beams for initial coarse acquisition and high-frequency narrow beams for refinement.
Design can integrate two-stage approach with the Gearbox-PHY architecture to manage control and data planes effectively, specifically utilizing it for $6$G devices like XR glasses to maximize their battery life.
The architecture requires balancing the low resolution of $8 \GHz$ arrays against the high precision but high losses of $24 \GHz$.}

\section{Near-Field \& Estimation Limits at FR3}
\label{sec:NF-FF}
\begin{figure}[t!]
    \centering
    \includegraphics[width=0.5\textwidth]{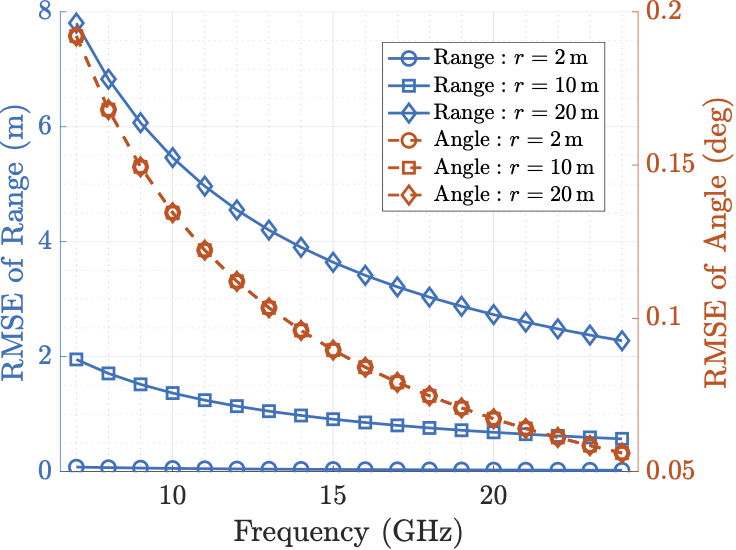}
    \caption{\textcolor{black}{Range (blue) and angle (orange) {CRB} versus carrier frequency for target distances of $2$m, $10$m, and $20$m under a $20\dB$ array-level {SNR}. The figure highlights how higher FR3 frequencies and nearer targets benefit from stronger {NF} effects, each at fixed {SNR}.}}
    \label{fig:NFFFCRB}
\end{figure}

\textcolor{black}{\ac{FR3} wideband affects range as targets transition between \ac{FF} and \ac{NF}. Array geometry matters; larger electrical apertures and wider bandwidths increase sensitivity to delay and direction, improving accuracy at a given \ac{SNR}. Broad \ac{FR3} bandwidth improves range discrimination and reduces multipath ambiguity. Angular discrimination improves with frequency and element count, allowing narrower mainlobes and lower sidelobes. In the \ac{NF}, array curvature makes coupled delay-angle estimation more informative.}

\textcolor{black}{We use the \ac{CRB} as a fundamental mathematical bound to establish the absolute limit on how precisely a target's range and angle can be estimated. We choose the \ac{CRB} because it represents the precision of any unbiased estimator attempting to jointly estimate the range and angle, which is equal to the reciprocal of the Fisher information.}Figure \ref{fig:NFFFCRB} shows how, at \ac{FR3} frequencies, the \ac{CRB} for range and angle estimation improves dramatically across three target distances ($2$m, $10$m, and $20$m). \textcolor{black}{For the \ac{CRB} evaluation, we model a $32$ element \textcolor{black}{\ac{ULA}} with half-wavelength spacing. We adopt a narrowband continuous-wave assumption to isolate the specific benefits of \ac{NF} wavefront curvature from wideband beam-squint effects.} 
\textcolor{black}{Furthermore, the \ac{CRB} for angle estimation is  independent of the target's range, as it relies on the constant slope of the phase of the incoming wave. In contrast, range estimation relies on the wave's spherical curvature. As the target distance increases, the spherical wavefront flattens and approaches \ac{FF} conditions, which makes the curvature harder to measure, hence degrading the range estimation accuracy.}Even at $20$m, where the source is closer to the \ac{FF} limit, higher frequencies still exhibit lower estimation errors. These results assume a $20\dB$ array-level \ac{SNR}, highlighting that geometric benefits from higher frequencies.
In essence, operating in the FR3 band provides a \ac{NF} advantage: it opens up more precise sensing opportunities by exploiting the richer propagation characteristics at those frequencies, making it highly attractive for short-to-mid-range radar and positioning applications.
\textcolor{black}{Operating on \ac{FR3} provides a  \ac{NF} advantage by exploiting spherical wavefront curvature to improve the fundamental precision of range estimation for short-to-mid-range applications.
Design can leverage the coupled delay-angle structure inherent in the \ac{FR3} \ac{NF} regime to improve delay resolution by using algorithms that acknowledge the \ac{NF} structure.
Closer distances and higher \ac{FR3} frequencies are good for range estimation when utilizing the steering vector.}

\section{FR3 Intra \& Inter Beam Squint Phenomena}
\label{sec:squint}
\textcolor{black}{Before detailing the specific phenomena observed in FR3, it is helpful to establish the concept of beam squint. In standard narrowband systems, antenna arrays steer beams toward a target by applying a specific phase shift to each antenna element. However, as 6G and ISAC push for wider bandwidths available in FR3, the physical wavelength of the signal varies significantly from the lower edge of the band to the upper edge. Because the hardware's phase shift remains fixed while the actual wavelength changes, the array radiation pattern can be frequency dependent, which leads to misalignment of beams, a phenomenon termed \emph{beam squinting} in the literature. 
\textcolor{black}{When operating across the bandwidths of the non-contiguous FR3 spectrum,}this frequency-dependent directional shift can be classified as: intra-beam and inter-beam squint.}
\begin{figure}[t!]
    \centering
    \includegraphics[width=0.5\textwidth]{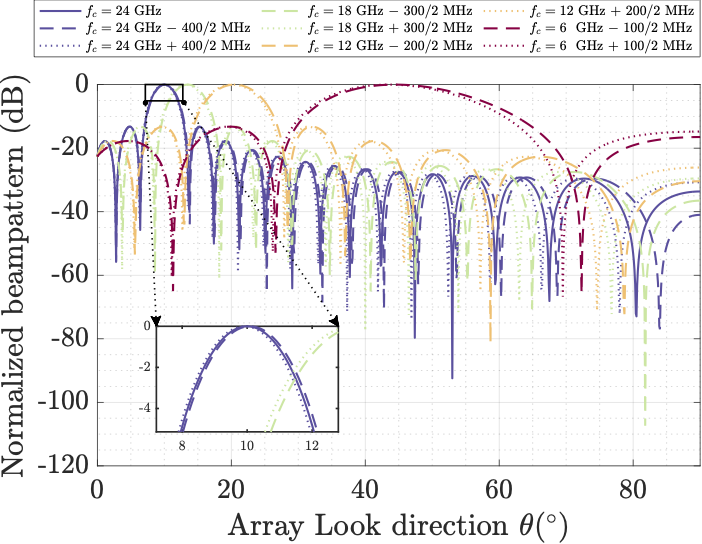}
   \caption{\textcolor{black}{Intra-beam squint and inter-beam squint effects occurring at FR3 bands.}}
    \label{fig:bmsqnt2}
\end{figure}

\subsection{Intra-beam squint effect}
\label{sec:intra-beam}
\textcolor{black}{The intra-beam squint effect refers to beam misalignments that occur within an \ac{FR3} band during beam coverage due to frequency differences across the band itself. The phenomenon is driven by a frequency-dependent radiation pattern, which causes different frequencies within the same beam to be radiated in slightly different directions. 
In Fig. \ref{fig:bmsqnt2}, consider a standard \textcolor{black}{\ac{ULA}} tasked with focusing on a specific target direction of $10^{\circ}$, where antenna spacing is optimized for $24 \GHz$ frequency.
When the array receives a signal at $10^{\circ}$ that spans a $400$ MHz bandwidth centered at $24 \GHz$, the combiner correctly peaks at the correct target direction. However, because the physical wavelength shifts slightly across that $400 \MHz$ band, the beam squints and deviates by $0.1^\circ$ at the bandwidth edges. 
\textcolor{black}{This deviation can increase for different FR3 frequencies, e.g., across a $100$ MHz bandwidth centered at $6 \GHz$ , we notice a deviation of $0.5^\circ$. 
It is important to clarify that while the $6$ GHz carrier undergoes a massive spatial shift relative to the $24$ GHz baseline (which is the inter-beam effect discussed next), the $0.5^\circ$ deviation  characterizes the intra-beam spreading across the $100 \MHz$ bandwidth itself.}


}

\subsection{Inter-beam squint effect}
\label{sec:inter-beam}
\textcolor{black}{
The inter-beam squint effect is a phenomenon that occurs between two or more beams at different FR3 carriers, resulting in the beams pointing in different directions, which causes beam misalignments that are much more dramatic than intra-beam squint, as using the same combiner creates significant directional deviations across different frequencies.

The inter-beam squint is tied to phase-shift of beamforming architectures. Because the \ac{FR3} spectrum covers a massive $\sim 4\times$ frequency span (from roughly $7 \GHz$ to $24 \GHz$), the electrical length of a fixed analog phase shift varies across the band. 
Consequently, when a base station aggregates non-contiguous carriers across this wide range, the beams for different carriers will point in different directions. For example, a signal arriving at $10^{\circ}$ covering 300 MHz centered around 18 GHz will peak around $13.4^{\circ}$ instead of $10^{\circ}$. We refer to this great deviation as the inter-beam squint effect. In \ac{mMIMO} deployments utilizing spatial multiplexing, inter-beam squint is a critical concern because it destroys the spatial isolation among wideband users, inevitably leading to severe inter-user interference because of the reduced spatial isolation between the users. 
}

\subsection{Solutions on \ac{FR3}}
As far as \ac{ISAC} is concerned, the inherent problem caused by FR3 \textcolor{black}{intra-} and inter-beam squint effects can deteriorate both communication rates, as well as sensing detection performance.
\textcolor{black}{To overcome the frequency-independence of standard phase shifters, \ac{FR3} beamformers require high flexibility. While \textit{multiband antennas} can mitigate squint by tailoring arrays to specific sub-bands, optimizing them for \ac{FR3} \ac{SnC} applications remains a key hardware challenge.}
A fully digital beamforming design can directly address the \textcolor{black}{intra-} and inter-squint problems by applying frequency-dependent phase shifts across the entire \ac{FR3} and sub-bands.
\textcolor{black}{Next, we evaluate how these \ac{FR3} acquisition and beamforming characteristics translate into sensing versus communication coverage.}
\textcolor{black}{\ac{FR3} suffers from intra-beam  that degrades target SNR, while attempting to reuse analog hardware across different FR3 sub-bands causes severe inter-beam pointing errors that can miss the target for sensing, and cause inter-user interference for communications.
\ac{ISAC} design must incorporate digital compensation techniques to  correct intra-/inter- beam squinting.
While leveraging the bandwidths available at higher \ac{FR3} frequencies is  necessary to achieve fine sensing resolution, doing so amplifies the squint penalty, forcing a tradeoff between analog hardware simplicity and the increased power required for digital beam correction.}

\section{ISAC Coverage on FR3}
\label{sec:isac-coverage}
\subsection{SER vs Sensing required SNRs: What QoS?}
\textcolor{black}{\ac{FR3} communication achieves a scalable tradeoff between coverage and achievable rate. Sensing coverage depends on the specific task (detection, estimation, or imaging) and associated QoS (detection/false-alarm probabilities). In target detection, sensing requires solving fewer hypothesis tests than communication symbol detection. Consequently, sensing demands lower \ac{SINR} and can achieve greater coverage than communication at  same power. Conversely, imaging requires estimating reflection coefficients for many targets, demanding higher \ac{SINR} and reducing coverage compared to communication.}

The above discussion highlights the need for an in-depth investigation of sensing coverage in the FR3 bands. Unlike the sparse propagation characteristics of mmWave frequencies, FR3 bands can provide richer multipath components, including both \ac{LoS} and \ac{NLoS} paths. Sensing \ac{QoS} may benefit from \textcolor{black}{\ac{NLoS} echo power}, potentially extending coverage beyond what is achievable in mmWave bands.
Given the pathloss and \textcolor{black}{\ac{QoS} targets, pilot-only sensing is insufficient at \ac{FR3}, making payload signals necessary.}

\subsection{Sensing With Data Payload Signals}
\label{sec:sensing-data-payload}
\textcolor{black}{Current \ac{ISAC} schemes rely on \ac{PHY} pilots, which occupy only $10$-$15$\% of time-frequency resources. The constrained allocation provides inadequate resolution and power for robust sensing, a problem significantly exacerbated by \ac{FR3}'s inherent path loss.}

\textcolor{black}{
\emph{The role of deterministic-random tradeoff:}
At this point, we encounter a fundamental physical dilemma known as the ``\ac{DRT}''. In essence, to achieve high communication rates, a waveform must carry as much  information as possible, which requires the transmitted signal to be "as random as possible" to maximize mutual information. Conversely, radar sensing prefers deterministic signals to achieve low delay-Doppler sidelobes and stable estimation performance. Consequently, when highly random data payloads are repurposed for sensing, the lack of guaranteed correlation properties causes a loss of sensing degrees of freedom, which degrades estimation performance compared to using deterministic pilots.
}
\textcolor{black}{So, to overcome the aforementioned challenges of integrating the remaining 85-90\% data \ac{PAS}, the communication signal must be optimized across baseband components:
\begin{itemize}
	\item \textbf{Constellation shaping}, whereby the statistical properties of constellation symbols impact sensing, e.g. the kurtosis of constellation symbols dictates the variance of the \ac{AF}. In particular, constellations with high kurtosis cause the \ac{AF} to fluctuate, whereas sub-Gaussian constellations with lower kurtosis (such as \ac{PSK}) generate more stable \acp{AF}. In fact, most modern communication systems use sub-Gaussian constellations. In order to jointly optimize \ac{SnC}, systems can employ probabilistic  constellation shaping to  control symbol fluctuations.
	\item \textbf{Orthonormal modulation basis}, where we know \ac{OFDM} minimizes ranging sidelobes under sub-Gaussian constellations, advanced modulation bases like \ac{OTFS} and \ac{AFDM} naturally exploit channel structures to deliver  stable parameter estimation in high-mobility ISAC scenarios.
	\item \textbf{Pulse shaping filters}, initially designed for communications to mitigate inter-symbol interference, pulse shaping filters can also shape the statistical properties of the \ac{AF} of random \ac{ISAC} signals.
\end{itemize}}
\textcolor{black}{Because pilot signals occupy only $10$-$15$\% of time-frequency resources, pilot-only allocations are insufficient; therefore, \ac{ISAC} systems must leverage the full data payload by optimizing baseband components as dictated by \ac{DRT}.
Design should optimize performance by employing probabilistic constellation shaping to control symbol fluctuations, leveraging advanced modulation bases, and utilizing pulse shaping filters to enhance the \ac{AF}.
While maximizing communication throughput requires random waveforms, reliable radar sensing demands deterministic signals for stable estimation, as described by \ac{DRT}.}

\section{\textcolor{black}{Case Study: Drone Localization}}
\label{sec:use-case}
\textcolor{black}{\begin{figure}[t!]
    \centering
    \includegraphics[width=0.5\textwidth]{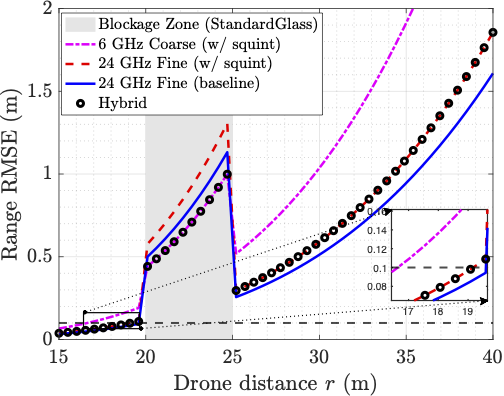}
   \caption{\textcolor{black}{Drone tracking coverage in the FR3 band: $6 \GHz$ coarse beam and the high-resolution $24 \GHz$ uncompensated beam.}}
    \label{fig:use-case}
\end{figure}
To illustrate the necessity of our proposed hierarchical alignment and intra-/inter- squint phenomenon, we evaluate an \ac{ISAC} scenario, namely tracking a low-airspace drone, which \textcolor{black}{presents} a sensing challenge due to their high mobility and small radar cross-section, demanding both high SNR, bandwidth, and integration gain to achieve reliable sub-meter localization. 
We evaluate the tracking \ac{RMSE} over distance under \ac{FR3} operating regimes, as illustrated in Fig. \ref{fig:use-case}.
When utilizing the lower $6 \GHz$ of the \ac{FR3}, the propagation enjoys favorable lower path loss relative to the mono-static \ac{ISAC} \ac{BS}. However, owing to the longer wavelength, the physical aperture of the array yields a relatively wide beam, as described in Section \ref{sec:beamalignment}. 
While efficient for the initial coarse scan, the lack of spatial resolution results in a higher \ac{RMSE}, e.g. $0.1 \meters$ accuracy at $16.6 \meters$.
Moving towards the upper end of the FR3 at $24 \GHz$ allows packing more antennas within the same aperture to combat the path-loss and  generate directive beams. The highly directive beams in the presence of in intra-beam squint can further achieve a $19.2 \meters$ coverage with $0.1 \meters$ accuracy with an additional $1 \meters$ accuracy if the squint is compensated. 
\textcolor{black}{Tracking a drone on \ac{FR3} reveals that while the lower $6 \GHz$ offers favorable path loss for coarse scanning, the $24 \GHz$ is needed for highly directive beams required for sub-meter localization.
Hybrid designs can make use of all \ac{FR3} bands, e.g. hierarchical alignment that leverages the $6 \GHz$ for initial acquisition and transitions to a squint-compensated $24 \GHz$ array to reliably extend the $0.1 \meters$ accuracy tracking coverage beyond $19.2 \meters$.
\textcolor{black}{During the standard glass blockage zone (in $20$–$25$ meters), the highly attenuated 24 GHz signal experiences severe penetration losses, causing its localization error to significantly increase.
The hybrid approach falls back to the robust 6 GHz frequency to maintain a lower \ac{RMSE}.

}

}

}

\section{FR3 Channel Simulators for ISAC \& Digital Twins}
\label{sec:channel-simulators}
\textcolor{black}{Deterministic modeling, such as \ac{RT}, is crucial for precise radio wave propagation simulation and channel state prediction, especially at higher frequencies where interactions with objects become more deterministic. Calibration with real-world data is essential, especially for FR3, where little empirical data exists.}
In the FR3 band, signal paths are shaped by penetration losses and interactions such as reflection, diffraction, whereas scattering may be ignored at FR2 bands, making calibration essential for realistic performance predictions, which is crucial for FR3 because all the \ac{AS}, \ac{DS}, diffraction, penetration losses are now frequency dependent within FR3, making \ac{RT} calibration a vital channel modeling step.
Indeed, this is important for precise beamforming study, and managing interference in dense $6$G environments. 
While NYURay has been calibrated for $28$, $73$, and $142 \GHz$, similar calibration can be done on \ac{FR3}.
\textcolor{black}{Meanwhile, ISAC RT can be relevant to the assessment of attainable \ac{SE} and localization accuracies over \ac{FR3}.}

\textcolor{black}{Calibrated \ac{FR3} models and \ac{RT} engines form wireless \acp{DT} which aims at creating virtual replicas using real-time measurements. \acp{DT}  allow \ac{AI} pipelines to validate beamforming, \ac{ISAC} processing, and handover policies without perturbing operational systems. For sensing, \acp{DT} allow systematic exploration of resolution and detection performance, which facilitates \ac{FR3} scenarios that are costly to test at scale.}
\textcolor{black}{Deterministic \ac{RT} models and their integration into \acp{DT} are essential for simulating \ac{FR3} networks, as propagation characteristics are highly frequency-dependent on \ac{FR3}.
\ac{RT} calibration with real-world measurements to build continuously updated digital twins is necessary, and determines the realism of underlying \acp{DT}.}

\section{Massive MIMO \& ISAC}
\label{sec:maMIMO}
We can see that \ac{mMIMO} imposes itself as a natural regime to combat pathloss and penetration losses. 
Besides enhancing \ac{SE}, \ac{mMIMO} has already shown success in \ac{FR3} bands, for instance, at $11 \GHz$ \cite{6884147}, and is yet to be explored on  FR3. 
\paragraph{\ac{mMIMO} meets \ac{ISAC}} \ac{mMIMO} is studied for communications, but its effects on \ac{FR3} remain partially un-explored. 
Besides the increased capacity, we know that in terms of reliability, the probability of a \textit{link outage} scales as $\operatorname{SNR}^{-N_tN_r}$ (point-to-point \ac{mMIMO}). 
A fundamental question to answer would be \textit{How many \ac{mMIMO} antennas do we need to support \ac{ISAC} applications on \ac{FR3} ?}
More specifically, \ac{RMT} has already been used to provide tight approximations for achievable communication rates, but \textit{can \ac{RMT} for \ac{mMIMO} provide joint \ac{SnC} bounds?} For \ac{FR3} multi-band sensing, one may consider probability of detection expressions, under fixed false alarm following, for e.g., Neyman-Pearson test. \ac{CRB}-rate expressions can also be considered in such a context.
The analysis can turn out to be fruitful for scenarios where the \textit{number of targets, number of users and the number of antennas grow infinitely large at the same speed}.
Besides fundamental \ac{ISAC} bounds for \ac{mMIMO} at \ac{FR3}, \textit{pencil-like beams} become an important element for accurate beamforming (e.g. $3-4^\circ$ for $4000$ antenna array elements as opposed to $13-14^\circ$ in $5$G), allowing high angular resolution, and enabling precise beamforming and spatial separation. While this is good news for communications, due to spatial interference mitigation, sensing can also benefit from the accuracy naturally given by the narrow pencils.
 
 \paragraph{Favorable \ac{FR3} channel hardening} 
The effects of small-scale fading across the FR3 bands can be eliminated, all thanks to \textit{channel hardening}.
More precisely, the small-scale fading averages out over the antenna array dimension, i.e. its variance decreases proportionally with the number of antenna elements, a ramification of the law of large numbers.
A question remains: \textit{How can channel hardening behave across the entirety of \ac{FR3} bands, when each band has different propagation properties ?}
\textcolor{black}{\ac{mMIMO} imposes itself as a necessary architecture to combat \ac{FR3} pathloss making use of massive antenna arrays to generate accurate beams that improve \ac{SnC}. 
\ac{RMT} can be a useful tool to derive fundamental \ac{SnC} bounds  to determine the antenna requirements as targets and users scale.
Deploying a massive number of antennas for the \ac{mMIMO} regime introduces a tradeoff in hardware complexity, and thereby power consumption.}

\textcolor{black}{While this paper focuses on \ac{FR3} and \ac{ISAC}, the evolution is also trending toward distributed configurations, such as coordinated multi-point and cell-free \ac{MIMO}, which can offer enhanced coverage and interference management for \ac{ISAC}.}




\section{Conclusions}
\textcolor{black}{The \ac{FR3} spectrum provides a natural habitat for $6$G \ac{ISAC} systems due to legacy \ac{FR2} radars which \textcolor{black}{connect} MAC-managed \ac{RaaS} architecture, where operators can form an initial bridge to network-native sensing. Moreover,  hierarchical beam alignment is needed to balance coverage and precision, and optimize data-payload waveforms to navigate the deterministic-random tradeoff in order to fully leverage the upper-midband's potential. Furthermore, by coupling these physical layer innovations with calibrated digital twins and \ac{mMIMO} architectures, networks can exploit spatial diversity and opportunistic illuminators. Ultimately, these elements capture a coherent path from today's coverage-oriented networks to future $6$G systems, where \ac{FR3} ISAC provides distributed perception and connectivity for mission-critical verticals like low-airspace safety and \ac{XR}.}

\section{\textcolor{black}{ACKNOWLEDGMENTS}}
This work is supported by Tamkeen under the Research Institute NYUAD grant CG017.

\bibliographystyle{IEEEtran}
\bibliography{refs}

\vfill

\end{document}